**Superconductivity Discovered In Niobium Polyhydride At High Pressures**

X. He [a1,2,3], C. L. Zhang[a1,2], Z. W. Li[a1,2], K. Lu[a1,2], S. J. Zhang[1], B. S. Min[1,2], J. Zhang[1], L.C. Shi[1,2], S. M. Feng[1], Q.Q. Liu[1], J. Song[1], X. C. Wang*[1,2], Y. Peng[1,2], L. H. Wang[4], V. B. Prakapenka[5], S. Chariton[5], H. Z. Liu[6], C. Q. Jin*[1,2,3]

[1] *Beijing National Laboratory for Condensed Matter Physics, Institute of Physics, Chinese Academy of Sciences, Beijing 100190, China*

[2] *School of Physical Sciences, University of Chinese Academy of Sciences, Beijing 100190, China*

[3] *Songshan Lake Materials Laboratory, Dongguan 523808, China*

[4] *Shanghai Advanced Research in Physical Sciences, Shanghai 201203, China*

[5] *Center for Advanced Radiations Sources, University of Chicago, Chicago, Illinois 60637, USA*

[6] *Center for High Pressure Science & Technology Advanced Research, Beijing 100094, China*

Niobium polyhydride was synthesized at high pressure and high temperature conditions by using diamond anvil cell combined with in situ high pressure laser heating techniques. High pressure electric transport experiments demonstrate that superconducting transition occurs with critical temperature($T_c$) 42 K at 187 GPa. The shift of Tc as function of external applied magnetic field is in consistent to the nature of superconductivity while the upper critical field at zero temperature $\mu_0 H_{c2}(0)$ is estimated to~16.8 Tesla while the GL coherent length ~57 Å is estimated. The structure investigation using synchrotron radiation implies that the observed superconductivity may come from F*m*-3*m* phase of $NbH_3$.

[a] Authors contributed equally;

[*] Corresponding authors: wangxiancheng@iphy.ac.cn; jin@iphy.ac.cn

## Introduction

Niobium element hosts superconductivity (SC) with the record critical temperature $T_c$ at ambient pressure among the elements in the period table [1]. For niobium hydride of $NbH_n$ at ambient pressure, the dissolution of hydrogen and its effect on the SC of Nb has been studied[2, 3]. $NbH_n$ has several phases dependent on hydrogen concentration[2]. All the hydrogen atoms occupy the tetrahedral interstitial sites (T sites) of face centered cubic (fcc) Nb lattice when $n < 2$. In the case of $x < 0.04$ and at room temperature, the T sites are randomly occupied by hydrogen for α phase $NbH_n$; while for β phase $NbH_n$ ($0.7 < n < 1.1$) the hydrogen atoms are ordered in chains along a [110] direction. These two phases coexist when the hydrogen concentration $n$ is between 0.04 and 0.7[2]. The next phase with higher $n$ is δ phase of $NbH_2$ where all the T sites are occupied. It was reported that the dissolution of hydrogen reduces the electron density of state (DOS) mainly contributed by 3$d$ orbital near the Fermi level and dramatically suppresses $T_c$ of Nb[3]. When $n$ in $NbH_n$ exceeds 0.7, $T_c$ is suppressed to less than 1.3 K[2].

Sulfur polyhydride of $SH_3$ was experimentally discovered to host SC with $T_c$ 203 K at 155 GPa after the theoretical predictions[4, 5]. Soon after the discovery, a series of binary polyhydride superconductors have been experimentally reported[6-21]. Besides $SH_3$, there are several other polyhydride superconductors with $T_c$ exceeding 200 K, including $LaH_{10}$ (250~260 K at 170-200 GPa)[6, 7], alkali earth hydride of $CaH_6$ (210~215 K at 160 ~172 GPa)[8, 9], $YH_9$ [10]. As for transition metals mainly the early transition metals polyhydrides are investigated, and the $T_c$ values of $ZrH_n$, $HfH_{14}$ and $TaH_3$ were experimentally reported to be about 71 K at 220 GPa[11], 83 K at 243 GPa[12] and 30 K at 197 GPa[13], respectively. For the polyhydride with the element in IVA and VA groups, $SnH_n$, $PH_n$ and $SbH_4$ have been experimentally reported to have SC with $T_c$ 70 K at 200 GPa[14], 103 K at 226 GPa[15] and 116 K at 184 GPa[16], respectively. Also, the summary of experimentally reported binary hydride superconductors can been seen in Table I, where only the highest $T_c$ value is listed for each element hydride and the related data are refered to the review paper[22], the references[5-19] and this work. For current extensively studied polyhydride superconductors at high pressures, hydrogen 1s orbital has significant contributions to the DOS near Fermi level and results in high temperature SC. Most of the metals of these polyhydride superconductors are the elements of IIA, VIA and IIIB groups of the period table[16-21, 23]. Beside the early studies of niobium hydride at ambient pressure, the stability and SC of $NbH_n$ at high pressure have also been investigated [24-26]. It was theoretically predicted that NbH and $NbH_2$ should be stable within 300 GPa while $NbH_3$ and $NbH_4$ with higher hydrogen concentration should become stable only above 50 and 300 GPa, respectively[24]. In addition, *I4/mmm* phase of $NbH_4$ was predicted to host SC with $T_c$ about 38-47 K at 300 GPa[24]. $NbH_3$ was confirmed by experiments to exist above 56 GPa and crystallize in a distorted bcc structure with a space group of *I*-43*d* [25]. However there is no SC reported by experiments for niobium polyhydrides with higher hydrogen concentration at high pressure. Here we report the synthesis of niobium polyhydride at high pressure. We found that the polyhydride sample exhibits SC with $T_c$ 42 K at 187 GPa. The results suggest that the SC likely arise from F*m*-3*m* phase of $NbH_3$.

# Experiments

The niobium polyhydride samples were synthesized at high pressure and high temperature conditions by using diamond anvil cell (DAC) high pressure techniques. The culet diameter of diamond anvils is 50 μm beveled to 300 μm. The prepressed gasket of T301 stainless was drilled with a hole at the center of 300 μm in diameter, before filled with insulating aluminum oxide mixed with epoxy resin. It was further pre-pressed and drilled with a hole of 40 μm in diameter to act as high pressure chamber for the specimen. Ammonia borane (AB) was filled into the chamber that plays the role of pressure transmitting medium and hydrogen source as well. Pt was deposited on the surface of the anvil culet with the thickness of 0.5 μm used as inner electrodes. Niobium foil (99.9%) with the size of 10 μm*10 μm in plane and 1 μm in thickness was stacked on the inner Pt electrodes. Then DAC was clamped and pressure was applied. The pressure was determined by measuring the Raman peak of diamond. The details are referred to the ATHENA procedure reported in Ref. [27].

The sample at high pressure was heated by a YAG laser in a continuous mode with 1064 nm wave length. The focused laser beam size was about 5 μm in diameter. The temperature was determined by fitting the black body irradiation spectra. After the synthesis the pressure was kept unchanged while the sample was applied to electric conductivity measurements at high pressure. A Van der Pauw method was employed and the applied electric current was 0.1 mA. The high pressure transport experiments are performed in a MagLab system, which can provide synergetic extreme environments with a temperature down to 1.5 K and a magnetic field up to 9 Tesla[28-30].

*In-situ* high pressure x ray diffraction (XRD) measurements were performed at 13 IDD of Advanced Photon Source at the Argonne National Laboratory. The x ray wave length $\lambda$ = 0.3344 Å and the diameter of the beam line is ~3 μm. Symmetric

DACs were used for the XRD experiments, and Rhenium was used as gasket. The diameter of high pressure sample is about 25 μm. The pressure was calibrated by the equation of state for rhenium as well as Pt which was put into the high pressure chamber as internal pressure marker. The XRD images are converted to one dimensional diffraction data with Dioptas[31].

## Results & Discussions

The samples were synthesized at 187 GPa. After the first laser heating (Run 1), the sample was applied for resistance measurements. By keeping the pressure unchanged the sample was heated for the second time (Run 2). Fig. 1 presents the temperature dependence of resistance $R(T)$ measured at 187 GPa for samples of Run 1 and Run 2. Both samples show two step transitions at about 42 K and 34 K, respectively. Zero resistance has been achieved at low temperature as shown in the lower inset of Fig. 1, suggesting the transitions are superconducting type. The multiple superconducting transitions should be caused by the hydrogen with different concentrations in the synthesized sample, which has been observed in other synthesized hydride superconductor[9, 17]. For Run 2 the first resistance drop is much larger and the zero resistance temperature shifts towards high temperature relative to Run 1, which imply the second heating procedure can significantly enhance the superconducting volume fraction, especially the high $T_c$ superconducting phase. The heating process at such high pressure of nearly two megabar level is very challenge. It often cause the broken of diamond anvils in the experiments before perform electric conductance measurements. The anvils become broken when it was heated for the Run 3. Hence we adopt experimental results obtained from Run 2 that is so far the best from many tests. To clearly demonstrate the onset superconducting $T_c$ the derivative of resistance over temperature for Run 2 was plotted in the upper inset of Fig. 1. From the upturn of the derivative curve, the $T_c^{onset}$ value is determined to be 42 K.

To investigate the SC under magnetic field, the temperature dependence of resistance measurements under different magnetic fields $H$ for sample Run 1 were performed as shown in Fig. 2(a). The superconducting transition is gradually suppressed by magnetic fields, in consistence with the nature properties of SC. The dashed line marks the resistance that the value is 95% of the normal state at $Tc^{onset}$, which is used as the criteria of $T_c^{95\%}$. Fig. 2(b) presents the critical field $H_{c2}$ versus temperature. The $T_c$ value is dramatically suppressed from 38.1 K to 31.7 K by a field of $H$ = 4 T. The upper critical magnetic field at zero temperature of $\mu_0 H_{c2}^{Orb}(0)$ controlled by orbital deparing mechanism in a dirty limit can be determined by the Werthamer-Helfand-Hohenberg (WHH) formula of

$$\mu_0 H_{c2}(T) = -0.69 \times dH_{c2}/dT\,|_{Tc} \times T_c,$$

where $dH_c/dT$ is the slope near $T_c$ with $H$ = 0 T. Since the $T_c$ at $H$ = 4 T is far away from that at zero field, only the $\mu_0 H_{c2}(T)$ data within $H$ = 1 T was linearly fitted as shown in the inset of Fig. 2(b), from which the slope of $dH_c/dT$ can be obtained to be about -0.52±0.02 T/K. By using the WHH formula and taking $T_c^{95\%}$ = 38.1 K, the $\mu_0 H_{c2}^{Orb}(0)$ value can be calculated to be ~13.6±0.5 T. Also, $\mu_0 H_{c2}(0)$ can be estimated by the Ginzburg Landau (GL) formula of

$$\mu_0 H_{c2}(T) = \mu_0 H_{c2}(0)(1 - (T/T_c)^2),$$

as shown in Fig. 2(b). The GL formula fitting leads to a $\mu_0 H_{c2}(0)$ value of ~13.1±0.3 T, which agrees well with that estimated by WHH method. In addition, the electron Zeeman energy dependent on high magnetic field can make a significant contribution to departing Cooper pair and determining $\mu_0 H_{c2}(0)$. For the case of weak coupling superconducting system, the $\mu_0 H_{c2}(0)$ limited by such a spin deparing mechanism is determined by the formula of $\mu_0 H_{c2}^{P}(0) = 1.86 \times T_c$. Thus, the $\mu_0 H_{c2}^{P}(0)$ for the niobium polyhydride superconductor can be calculated to be 70 T, which is much larger than $\mu_0 H_{c2}^{Orb}(0)$. This implies that here the Cooper pair is broken through the orbital depairing mechanism. According to the equation of $\mu_0 H_{c2}(0)= \Phi_0/2\pi\xi^2$, where $\Phi_0= 2.067 \times 10^{-15}$ Web is the magnetic flux quantum, the GL coherent length $\xi$ is calculated to be ~50.3 Å by taking $\mu_0 H_{c2}(0) = 13$ T.

*In situ* high pressure *x* ray diffraction experiments were performed to study the superconducting phase. Another sample was synthesized under 184 GPa with sufficient heat to generate the high $T_c$ superconducting phase. Fig. 3(a) displays the *x*-ray diffraction pattern collected with the synthesized pressure unchanged. Except for the diffraction peaks from Re gasket, all the other peaks can be indexed on the basis of a cubic lattice with a space group of F*m*-3*m*. Previous works[25] have experimentally demonstrated that F*m*-3*m* phase of $NbH_2$ only occurs at low pressure with hydrogen atoms occupying the T sites of fcc Nb lattice, and it reacts with $H_2$ to form $NbH_{2.5}$ with a hexagonal structure at about 39 GPa. With further compression the I-43*d* phase of $NbH_3$ appears with a distorted body centered cubic (bcc) structure above 56 GPa[25], which is theoretically predicted to be stable up to 287 GPa[24]. For the group VB element of Ta, its polyhydride of $TaH_3$ has been experimentally reported to have such a distorted bcc structure. However, in this work for niobium polyhydride samples, fcc structure is found to be stable at the pressure of 180~190 GPa. Therefore we use the F*m*-3*m* phase of Nb as the initial structure model to carry out the refinement for our diffraction. The refinements smoothly converge to $R_{wp}$ =

2.1% and $R_p$ = 1.3%, respectively. The refined parameters of $a$ is about 4.0936 Å. Considering $NbH_4$ with higher hydrogen concentration can only exist above 287 GPa, we proposed that the observed fcc phase at 184 GPa should be F$m$-3$m$ phase of $NbH_3$, which would be transformed from the distorted bcc phase of $NbH_3$. The schematic view of the fcc structure of $NbH_3$ is shown in Fig. 3(b). The Nb atoms are located at the fixed 4$a$ Wyckoff positions of (0, 0, 0) and hydrogen atoms are sited at the fixed 8c and 4b Wyckoff positions of (0.25, 0.25, 0.25) and (0.5, 0.5, 0.5), respectively. For the F$m$-3$m$ phase of $NbH_3$, both the tetrahedral and octahedral sites of fcc Nb lattice are occupied by hydrogen atoms. There exist $H_{14}$ cages in the polyhydride with the neighbor hydrogen distance H~H about 1.77 Å. It seems that for H 1s electrons in such a distance can support three dimensional conducting path while further to superconductivity in F$m$-3$m$ phase of $NbH_3$. In addition, it has been experimentally reported that the dhcp phase of $NbH_{2.5}$ can coexist with $I$-43$d$ phase of $NbH_3$ above 56 GPa[25]. Therefore, it is proposed that some kind of phase of $NbH_{3-x}$ with partially O-site occupied might also exist in our sample under 187 GPa and correspond to the second superconducting transition. The details of the SC scenario in niobium polyhydide deserves theoretical studies in the future.

# Conclusion

Niobium polyhydride was successfully synthesized at high pressure and high temperature conditions. The SC with $T_c$ about 42 K was observed and the upper critical magnetic field $\mu_0 H_{c2}(0)$ is estimated to be 16.8 Tesla. A niobium polyhydride with F$m$-3$m$ phase was observed at 184 GPa that is proposed to be $NbH_3$ responsible to the observed SC.

Table I Periodic table of experimentally reported binary hydride superconductors under high pressure. Only the highest $T_c$ value is listed for each element hydride. The related data were refered to the review papers [22]and the reference[16] and this work.

Legend: $SH_3$ / 203 / 155 — Hydride / $Tc$ (K) / Pressure (GPa)

| IA | IIA | IIIB | IVB | VB | VIB | VIIB | | VIIIB | | IB | IIB | IIIA | IVA | VA | VIA | VIIA | VIIIA |
|---|---|---|---|---|---|---|---|---|---|---|---|---|---|---|---|---|---|
| H | | | | | | | | | | | | | | | | | He |
| Li | Be | | | | | | | | | | | B | C | N | O | F | Ne |
| Na | Mg | | | | | | | | | | | Al | $SiH_4$ 17 96 | $PH_3$ 103 207 | $SH_3$ 203 155 | Cl | Ar |
| K | $CaH_6$ 215 170 | $ScH_3$ 18.5 131 | Ti | V | Cr | Mn | Fe | Co | Ni | Cu | Zn | Ga | Ge | As | Se | Br | Kr |
| Rb | Sr | $YH_9$ 262 182 | $ZrH_6$ 71 220 | $NbH_3$ 42 180 GP | Mo | Tc | Ru | Rh | $PdH_x$ 9 0 | Ag | Cd | In | $SnH_{12}$ 70 200 | $SbH_4$ 116 184 | Te | I | Xe |
| Cs | $BaH_{12}$ 20 140 | $LaH_{10}$ 260 180 | $HfH_{14}$ 83 240 | $TaH_3$ 30 190 | W | Re | Os | Ir | PtH 6.7 30 | Au | Hg | Tl | Pb | Bi | Po | At | Rn |
| Fr | Ra | Ac | | | | | | | | | | | | | | | |

| | | | | | | | | | | | | | |
|---|---|---|---|---|---|---|---|---|---|---|---|---|---|
| $CeH_{10}$ 115 95 | $PrH_9$ 9 126 | $NdH_9$ 4.5 126 | Pm | Sm | Eu | Gd | Tb | Dy | Ho | Er | Tm | Yb | $Lu_4H_{23}$ 71 218 |
| $ThH_{10}$ 161 174 | Pa | U | Np | Pu | Am | Cm | Bk | Cf | Es | Fm | Md | No | Lr |

# Figure Captions

**Figure 1**. (a) Temperature dependence of resistance for samples of Run 1 & Run 2 measured at 187 GPa. The lower inset is the enlarged view of resistance curve, showing zero resistance was achieved in the SC states. The upper inset is the derivative of resistance over temperature for Run 2 to show the critical temperature of superconducting transition.

**Figure 2**. (a) Superconducting transition for sample Run 1 measured at 187 GPa and different magnetic fields. (b) The upper critical magnetic field $\mu_0 H_{c2}$(T) versus temperature. The red line is the fitting via GL theory. The inset shows the linear fitting for the $\mu_0 H_{c2}$(T) within $H$ = 1 T. The red star marks the $\mu_0 H_{c2}$(T) determined by WHH method.

**Figure 3**. (a) The *x* ray diffraction pattern collected under 184 GPa and the refinement. (b) The schematic view of the structure of F*m*-3*m* phase of $NbH_3$ and the $H_{14}$ cages. The blue and light pink balls are the hydrogen atoms, occupying the octahedral and tetrahedral sites of fcc Nb structure, respectively. The olive balls denote the Nb atoms.

# Fig. 1

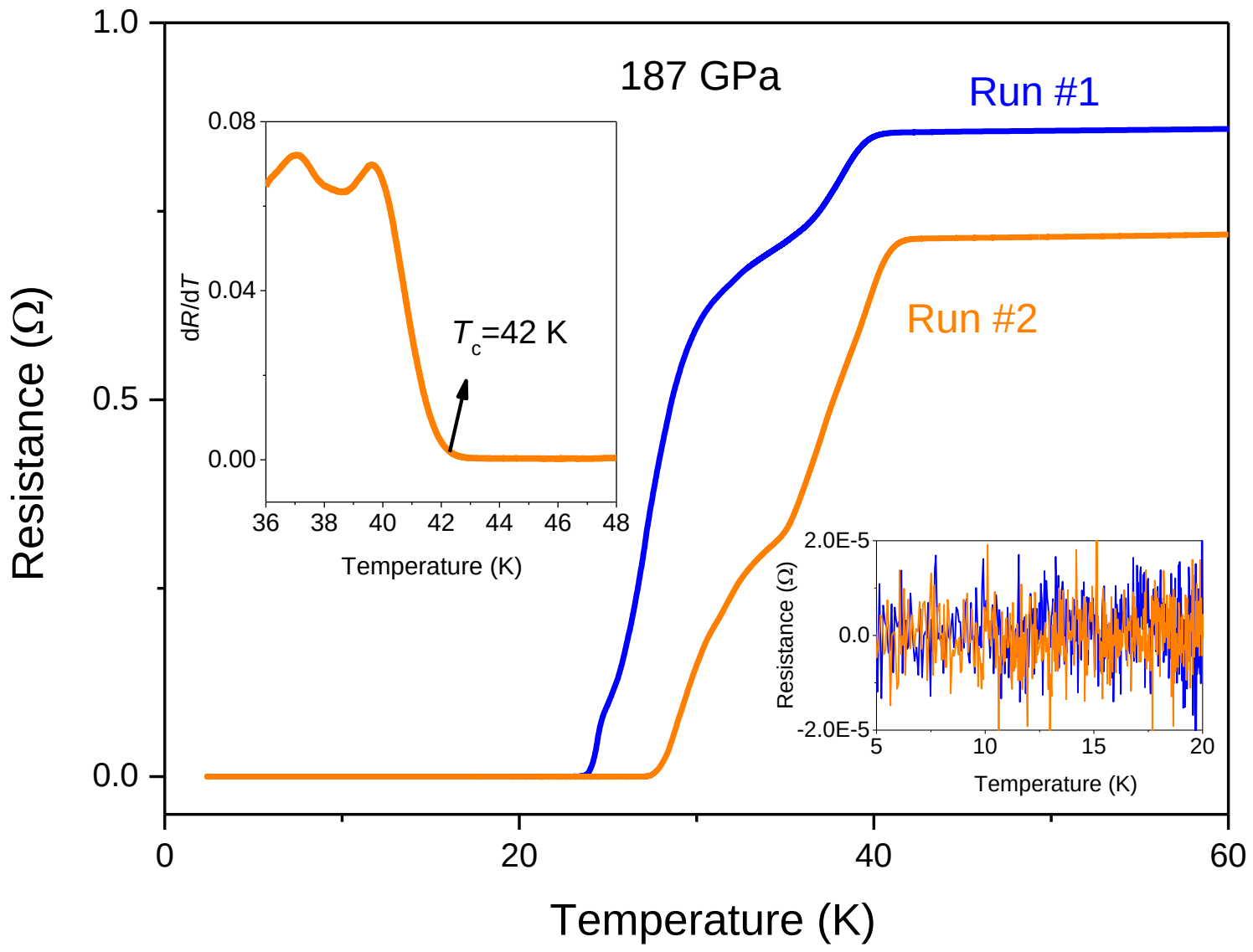

# Fig. 2 (a, b)

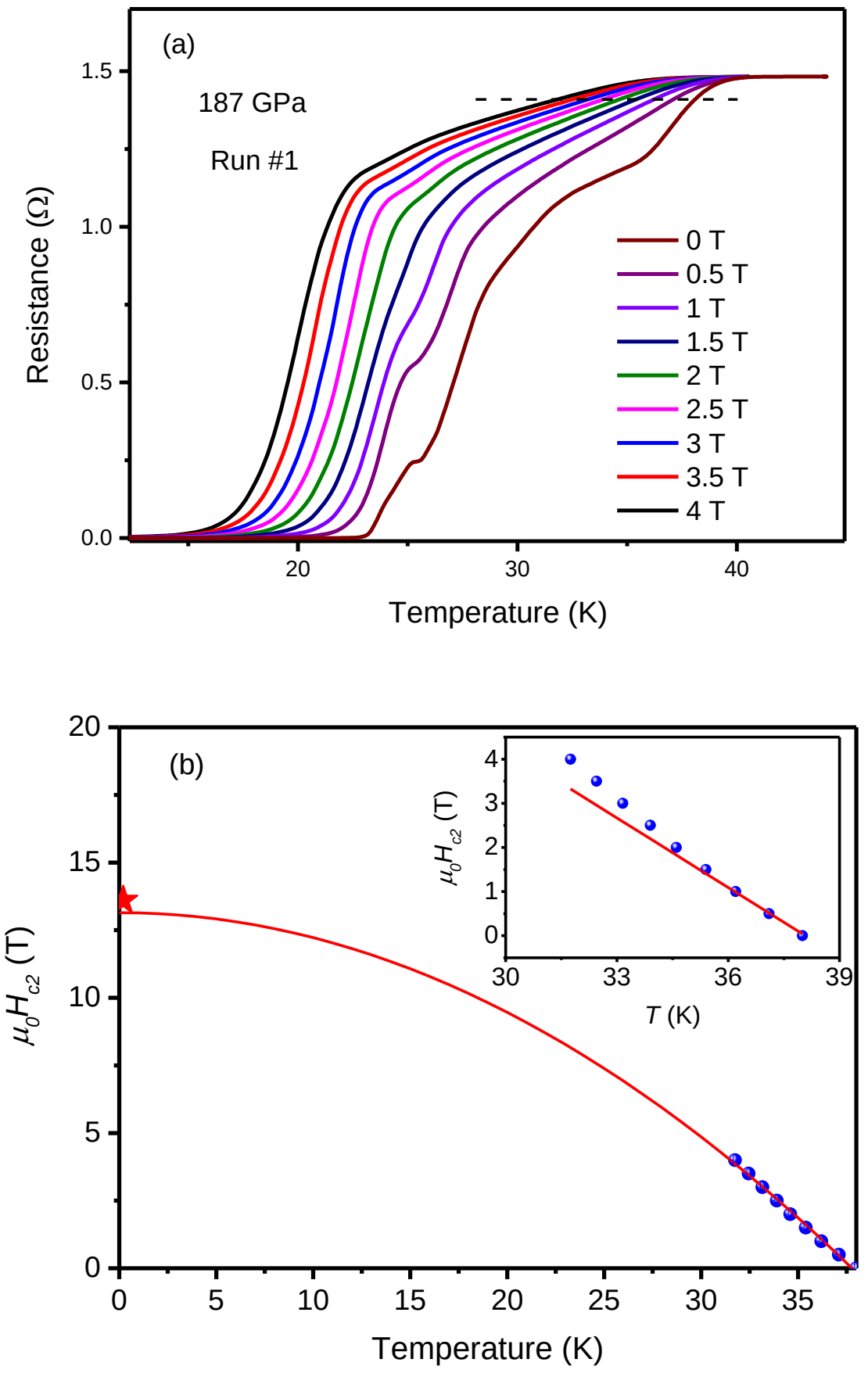

# Fig. 3

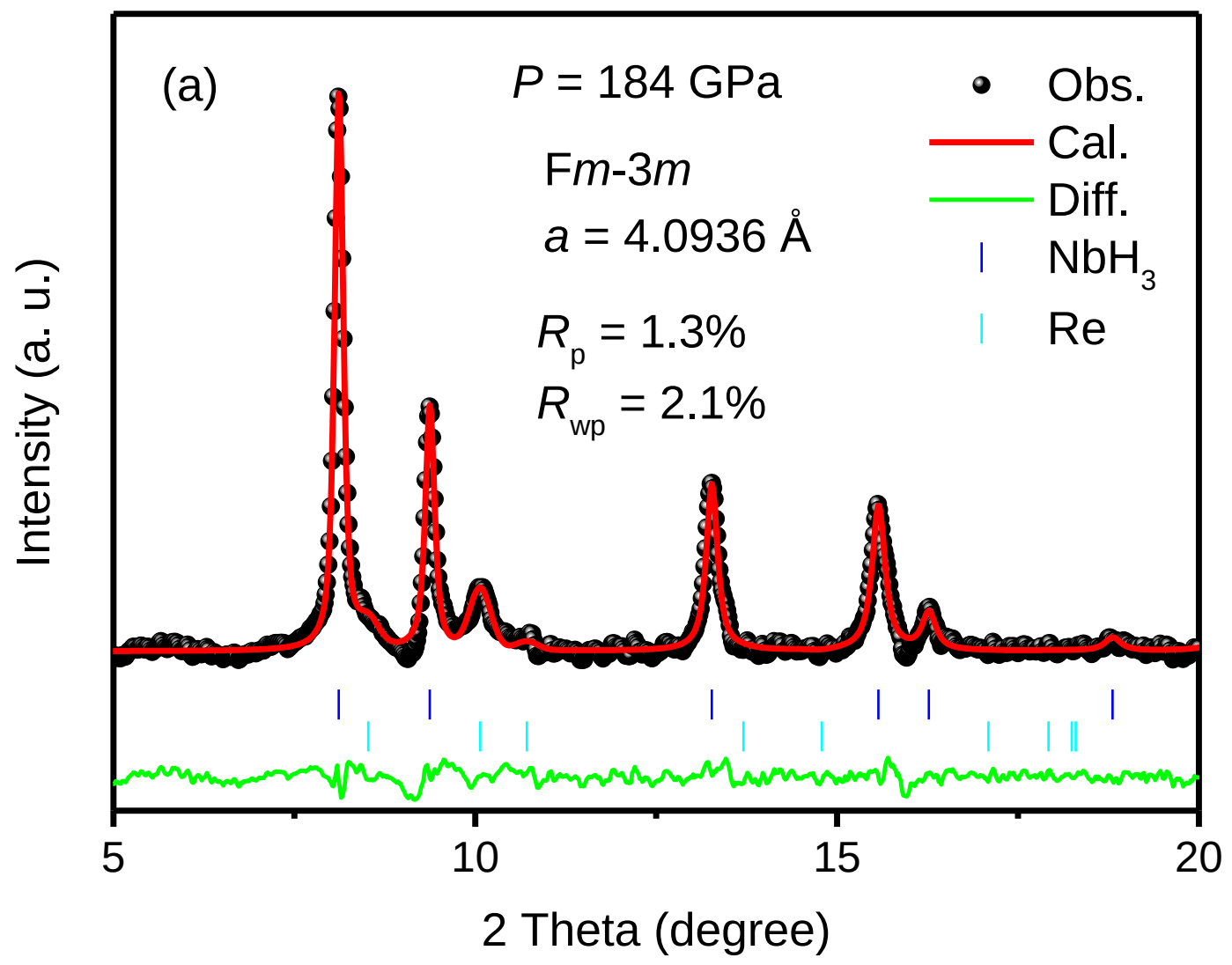

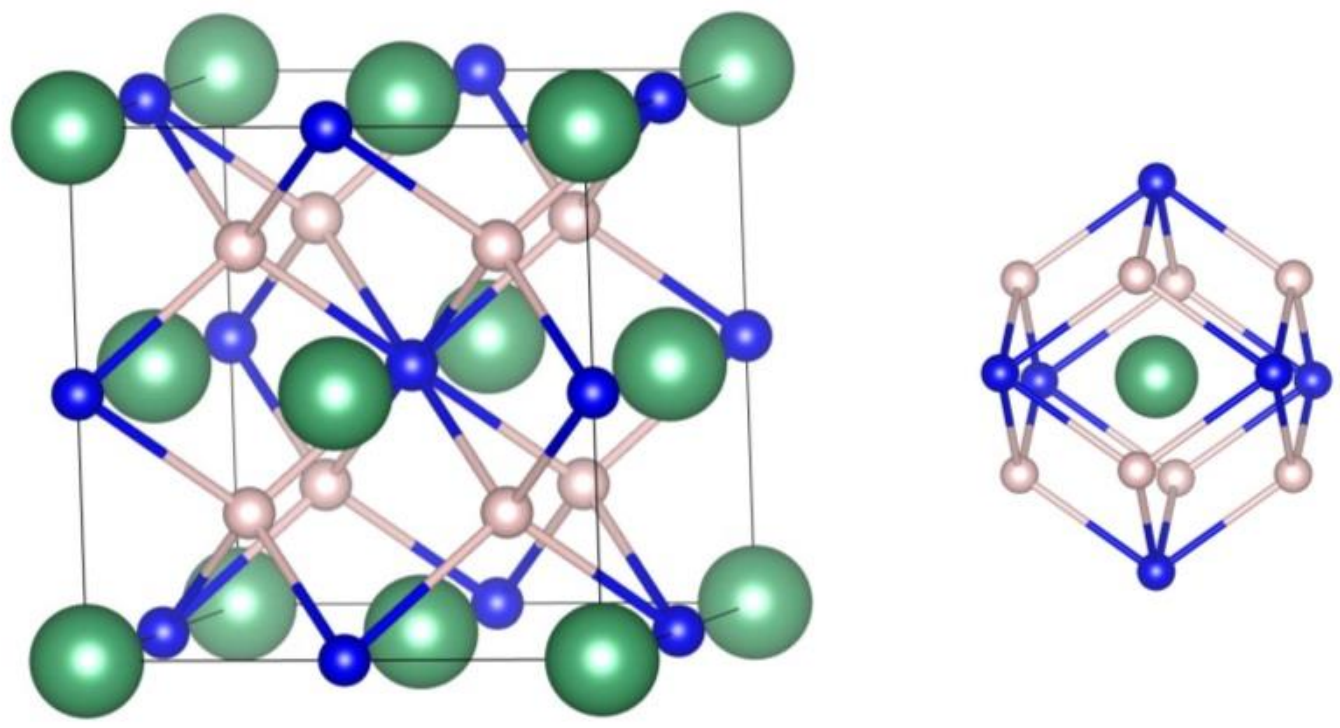

(b)